\documentclass{article}
\usepackage[utf8]{inputenc}
\usepackage{amsmath,amssymb}
\usepackage{graphicx}
\usepackage[a4paper,top=3cm,bottom=2cm,left=3cm,right=3cm,marginparwidth=1.75cm]{geometry}

\title{On the energy of the protons producing \\ the very high-energy astrophysical neutrinos}
\author{ Esteban Roulet$^{(a)}$ and Francesco Vissani$^{(b,c)}$\\
$^{(a)}$Centro At\'omico Bariloche, Comisi\'on Nacional de Energ\'\i a At\'omica\\ 
Consejo Nacional de Investigaciones Cient\'\i ficas y T\'ecnicas (CONICET)\\ 
Av. Bustillo 9500, R8402AGP, Bariloche, Argentina\\
$^{(b)}$ INFN, Laboratori Nazionali del Gran Sasso, 67100 Assergi, L’Aquila, Italy\\ 
$^{(c)}$ Gran Sasso Science Institute (GSSI), Viale F. Crispi 7, 67100 L’Aquila, Italy}
\date{}

\begin{document}

\maketitle

\begin{abstract}
We study the distribution of the energy of the protons that can produce an astrophysical  neutrino with a given observed energy, in the TeV--PeV range, both through $pp$ or $p\gamma$ interactions. Due to the increasing multiplicity of the pion production at high center of mass energies, the resulting average proton energies can be much larger than the often used approximate value $E_p\simeq 20E_\nu$. Also the threshold of the $p\gamma$ process can lead to a pronounced  increase in the values of $\langle E_p\rangle /E_\nu$ for decreasing neutrino energies. The results depend  sensitively on the assumed proton spectrum, since steeper spectra give less weight to the lower energy neutrino tail resulting from the decays of the abundant low-energy pions. In the $p\gamma$ scenarios they also depend sensitively on the spectrum of the target photons. The results are in particular relevant to  relate  possible characteristics of the neutrino spectrum to those of the corresponding cosmic rays that generated them. We also discuss the associated production of gamma rays at the sources.

\end{abstract}

\section{Introduction}

Very high energy ($>10$~TeV) neutrinos of non-atmospheric origin have been observed in the last decade with the IceCube detector \cite{ic,ic2,ic3,ic6,ic4,ic5}. They are believed to be produced in the astrophysical sources where the ultra-high energy cosmic rays (CRs) are accelerated, through the interactions of these with the gas or the radiation present in the acceleration sites or their surroundings. These interactions can produce unstable particles, mostly pions, which through their weak decays give rise to neutrinos, with  the decays of the neutral pions giving rise to comparable amounts of gamma rays (for recent reviews see \cite{me17,ah19}). Also the interactions of the CRs with the background radiation that they traverse as they propagate through the Universe, such as the cosmic microwave background (CMB) and the extragalactic background light (EBL), can give rise to high-energy cosmogenic neutrinos \cite{bz}. Cosmogenic neutrinos  are however likely subdominant with respect to the astrophysical ones produced at the sources, except eventually at energies well beyond 10~PeV \cite{ro12}. In spite of the major progress being done, the sources of the observed neutrinos are however still largely uncertain, with only one or two potential identifications having been done up to now through the coincidence between the arrival directions of well determined muon neutrino induced tracks and the location of the blazar TXS-0506+056 \cite{ictxs} or the Seyfert active galaxy NGC~1068 \cite{ic3}. The lack of correlations with source candidates of the remaining several tens of astrophysical neutrinos is also a relevant piece of the puzzle.

Besides the searches for angular correlations, also the detailed study of the energy distribution of the neutrinos should give important information about the mechanism producing the neutrinos, and hence on the nature of the CR sources. However, analyses using different neutrino datasets (muon tracks \cite{ic5}, cascade events \cite{ic6} or high energy starting events \cite{ic4}) don't lead to a definite picture about the astrophysical neutrino spectral shape. In particular, the slope $\gamma$ of a power-law fit $\phi_\nu\propto E_\nu^{-\gamma}$ which is obtained from the different studies actually covering the range of values between 2 and 3, so that more data would be required in order to clarify this issue. 

The two main classes of production mechanisms for the astrophysical neutrinos are the $pp$ and the $p\gamma$ interactions. The $pp$ mechanism takes place when a CR proton, or more in general a nucleon in a CR nucleus, interacts with the gas in the acceleration region (such as when a jet punches through the envelope of a star) or in the surrounding medium (as for instance in cluster of galaxies or starburst galaxies), while the $p\gamma$ one when it interacts with dense radiation fields, such as the synchrotron radiation from co-accelerated electrons in the jets of gamma ray bursts (GRB) or active galactic nuclei (AGN)  or the strong UV emission from the supermassive black-hole accretion disks in AGNs, etc.

Due to the approximate scaling invariance of the hadronic interactions, which implies for instance that the pion production yield of the $pp$ interactions mostly depends on the fraction of the proton energy carried by the pion, $x_\pi\equiv E_\pi/E_p$, rather than on the two energies separately, together with the small (logarithmic) energy dependence of the $pp$ cross section and the low threshold for pion production, one expects that in the $pp$ mechanism the spectrum of the neutrinos from the pion decays will have a spectral shape following closely that of the parent CR protons. In particular, a CR power-law differential spectrum $\phi_{\rm CR}(E_p)\sim E_p^{-\alpha}$ should approximately lead to $\phi_\nu(E)\sim E^{-\gamma}$, with $\gamma\simeq \alpha$. On the other hand, for the $p\gamma$ mechanism the proton energy threshold for the pion production to be possible is high, requiring that $E_p>m_pm_\pi(1+m_\pi/2m_p)/(2\varepsilon)\simeq 70\,{\rm TeV}/(\varepsilon/{\rm keV})$, with $\varepsilon$ being the target photon energy. In addition, at least near the threshold where the $\Delta$ resonant production dominates, the cross section has a significant energy dependence. As a result, the resulting neutrino spectrum depends not only on the spectral shape of the CRs, but also on the background photon spectrum.

In order to relate the energy of the neutrinos with that of the parent protons, one usually considers that the typical pion energy is $E_\pi\simeq E_p/5$ and that the four particles resulting from the decays $\pi^{+}\to \mu^{+}\nu_\mu$ and $\mu^{+}\to e^{+}\nu_e\bar\nu_\mu$ (and similarly for $\pi^{-}$) share similar amounts of energy, so that $E_\nu\simeq E_\pi/4\simeq E_p/20$. For the neutral pions one has that $\pi^0\to\gamma\gamma$, so that one similarly finds the naive estimate $E_\gamma\simeq E_p/10$.
These estimates are however very simplistic, since the pions have actually a distribution of energies  and similarly the energies of the neutrinos from muon decays are continuously distributed, and hence the neutrinos resulting from the interaction of a proton with a given energy are expected to have a quite broad distribution of energies. Moreover, the inelasticity of the interaction and the multiplicity of the produced particles depend on energy, with the multipion production becoming relevant at high energies and giving rise to large numbers of lower energy pions. The impact  of the neutrinos carrying different fractions of the proton energy in the final neutrino spectrum will also depend on the shape of the CR spectrum, since steeper (i.e. softer) spectra will make less relevant  the contribution from the low energy tail of the neutrinos from pion decays, enhancing the relative contribution of those with large values of $x_\nu=E_\nu/E_p$. On the other hand, flatter (i.e. harder) CR spectra  will lead to neutrino fluxes that depend more sensitively on the low energy tails and on the multipion production processes. The relevance of multipion production channels for the astrophysical neutrino production was emphasized e.g. in \cite{ke06} for the $pp$ scenarios, and in \cite{mu00,muna} for the $p\gamma$ scenarios.

In this work we will analyze  the distribution of the energies of the  protons that could have produced  neutrinos  with a given observed energy (in the TeV--PeV range). This is in particular useful to understand what are the CR energies being probed with the observed astrophysical neutrinos. We will also discuss the resulting shape of the spectra of the neutrinos for the different scenarios.  We will base the study on the analytical fits to the neutrino yields that parametrize the results of numerical simulations  obtained in \cite{ke06}, based on the SIBYLL 2.1 model \cite{sibyll} to account for the $pp$ hadronic interactions, and in \cite{ke08}, relying on the SOPHIA code \cite{sophia} to account for the $p\gamma$ interactions. Further studies of the photon spectra in $pp$ interactions were performed in \cite{ka14}, and the neutrino and photon spectra resulting in some post-LHC models were considered in \cite{su16}, obtaining results which are essentially consistent with those  we will adopt from \cite{ke06,ke08}.

\section{The $pp$ scenarios}
Let us consider the interactions of a flux of protons with energy $E_p$  with a target of hydrogen gas with density $n_{\rm H}$ (heavier target nuclei can approximately be accounted for by associating to them an effective H density, while for the generalization to the case of a flux of heavy nuclei see e.g. \cite{ka14,jo14}). The neutrinos produced can be computed as
\begin{equation}
    \frac{{\rm d}N_{\nu_i}}{{\rm d}E_\nu}=n_{\rm H}D\int_{E_\nu}^\infty \frac{{\rm d}E_p}{E_p}\sigma_{pp}F_{\nu_i}(x_\nu,E_p)\frac{{\rm d}N_p}{{\rm d}E_p},
    \label{dnude}
\end{equation}
where $\sigma_{pp}$ is the inelastic $pp$ cross section, with an approximate expression for its energy dependence  being $\sigma_{pp}\simeq (34.9+1.985L+0.18L^2)$~mb, with $L={\rm ln}(E_p/{\rm TeV})$  \cite{ka14}. The quantity $D$ is the distance traversed by the protons through the target gas, so that $N_{\rm H}\equiv n_{\rm H}D$ is the associated column density, and we have assumed that the source medium is thin, i.e. that $\sigma_{pp}N_{\rm H}\ll 1$. Given that $\sigma_{pp}\sim 10^{-25}$~cm$^2$, the source will be thin as long as $N_{\rm H}\ll 10^{25}$~cm$^{-2}$.  The function $F_{\nu_i}$, depending on $x_\nu\equiv E_\nu/E_p$ and on $E_p$, describes the neutrino yield in the $pp$ interaction, and is in principle different for the different neutrino and antineutrino flavors. At low energies there is a noticeable excess of positive pions being produced in $pp$ interactions, leading in particular to a net excess of $\nu_e$ over $\bar\nu_e$. However, when considering very high energies it is a good approximation to assume that the $\pi^{+}$ and $\pi^{-}$ multiplicities are similar. Hence, for the energies that we will explore here the fluxes of neutrinos and antineutrinos are expected  to be similar for the $pp$ scenarios, so that we will  consider for  each flavor the sum of neutrinos and antineutrinos that were obtained in \cite{ke06}, and just refer to them as neutrinos.  Note also that neutrino telescopes, such as IceCube, cannot distinguish neutrinos from antineutrinos, with the rare exception of the $\bar\nu_e$ induced $W$-resonance events at 6.3~PeV neutrino energies resulting from the interaction with atomic electrons ($\bar\nu_e e\to W\to anything$). The index $i$ in Eq.~(\ref{dnude}) identifies the type of neutrino produced, with $\nu_\mu^{(1)}$ being those from the pion decay, while $\nu_\mu^{(2)}$ and $\nu_e$ those from the subsequent muon decay. 
The spectrum of $\nu_e$ turns out to be, to an accuracy better than 5\%, equal to that of $\nu_\mu^{(2)}$, and hence the two fluxes are just assumed to be the same. To compute these functions one has to account for the fact that the muon produced in the pion decay is polarized \cite{lipari}, since this affects the energy distribution of its decay products. 
  These functions have been parameterized 
 in \cite{ke06} using simulations based on the SIBYLL~2.1 code to account for the details of the $pp$ hadronic interactions (finding comparable results also for simulations based on the QGSJet hadronic interaction model). There is also a small contribution to the neutrino fluxes from charged kaon decays  (which is below the 10\% level, see e.g. \cite{vi08}), but following \cite{ke06} we do not include it since this should not affect significantly our results.
 
\begin{figure}[t]
\centering
\includegraphics[scale=1,angle=0]{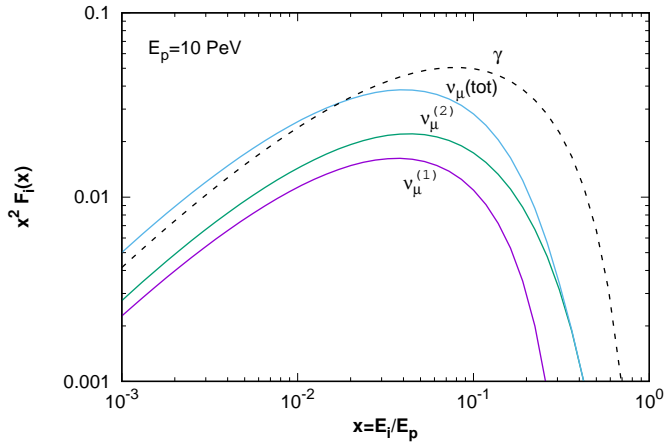}
\caption{Yields of neutrinos and photons, $x^2F_i(x,E_p)$ vs. $x$, for an incident proton energy $E_p=10$~PeV. The spectrum of $\nu_e$ is very similar to that of the  $\nu_\mu^{(2)}$ produced in the $\mu$ decays.}
\label{fvsx}
\end{figure}

 Analogous expressions also hold for the flux of secondary $\gamma$ rays, which arises mostly from $\pi^0$ decays, with a smaller contribution from the isosinglet $\eta$ decays. The different functions $F_i$ are plotted in Fig.~\ref{fvsx} for the illustrative case with $E_p=10$~PeV.
  
\begin{figure}[t]
\centering
\includegraphics[scale=1,angle=0]{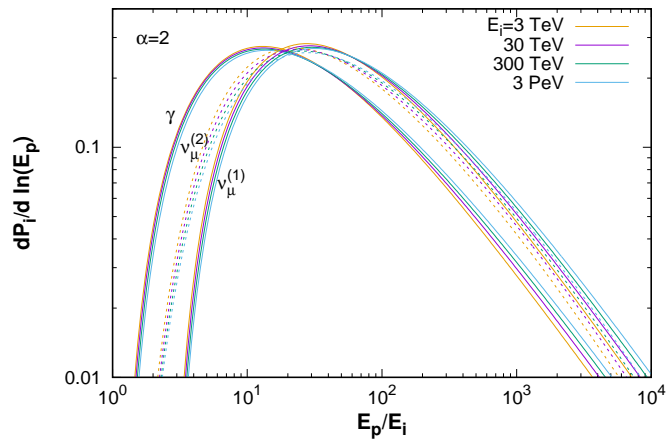}
\caption{Probability distribution for the parent proton energies that produce neutrinos or photons with different energies in $pp$ interactions, adopting $\alpha=2$. }
\label{pivsep}
\end{figure}

 The probability that a neutrino  of type $i$ (or a photon) and energy $E_i$  originated from protons in a given logarithmic energy interval is then
 \begin{equation}
     \frac{{\rm d}P_i}{{\rm d\,ln}E_p}(E_i,E_p)={\cal N}\sigma_{pp}F_i(x_i,E_p)\frac{{\rm d}N_p}{{\rm d}E_p},
 \end{equation}
 where the normalization is obtained from ${\cal N}^{-1}=\int_{E_\nu}^\infty \sigma_{pp}F_i [{\rm d}N_p/{\rm d}E_p]\,{\rm d}E_p/E_p$.
The resulting distributions are shown in Fig.~\ref{pivsep}, adopting a power-law CR spectrum with spectral index $\alpha=2$ and no cutoff. One can see that the distributions peak at different values for the different types of secondaries, they are quite broad and have significant high-energy tails.

\begin{figure}[t]
\centering
\includegraphics[scale=1,angle=0]{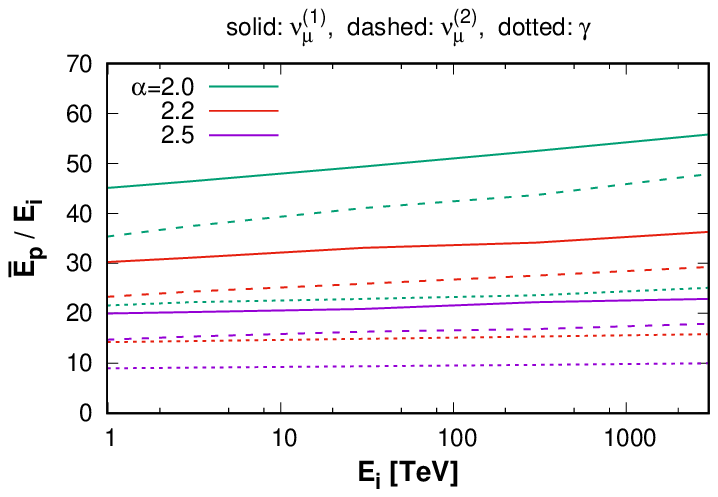}\includegraphics[scale=1,angle=0]{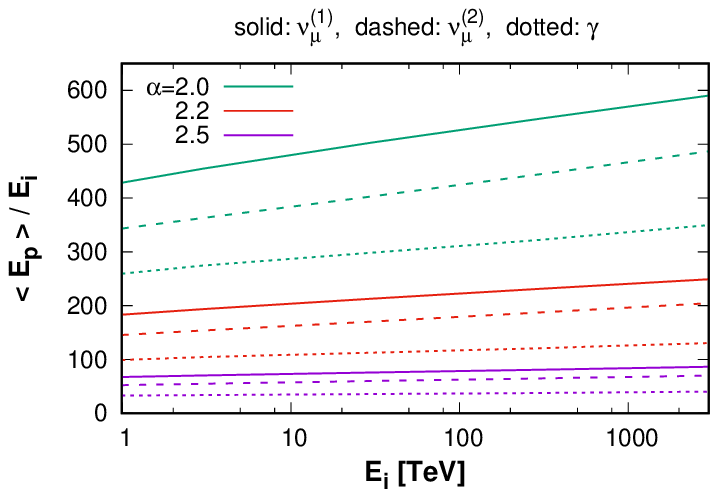}
\caption{Left panel: Ratio $\bar E_p/E_i$ involving the median proton energy $\bar E_p$ giving rise to a neutrino (or gamma) energy $E_i$, for different CR spectral indices $\alpha$, as a function of $E_i$. Right panel: similar plot but considering the average proton energy.}
\label{emed}
\end{figure}

A useful quantity is the median energy of the  protons, $\bar{E}_p$, leading to a given neutrino energy, which is defined such that the integral of the probability up to $\bar{E}_p$ gives 0.5. In the left panel of Fig.~\ref{emed} we plot the value of $\bar{E}_p/E_i$ for different representative values of  the neutrino energies and for different values of the spectral index $\alpha$ of the assumed power-law proton distribution. We see that the ratio  $\bar{E}_p/E_i$ depends on the kind of neutrino considered and on the value of $\alpha$. Its value increases with increasing neutrino energy,  due to the larger contribution from multipion production channels at higher energies. The values obtained differ from the usual assumption that $E_p/E_\nu\simeq 20$, specially when the proton spectrum is hard since in this case the contribution from neutrinos with small values of $x_\nu$ becomes more relevant. In particular, we see that for $\alpha\simeq 2$ a neutrino of 1~PeV will originate from protons with $\bar{E}_p\simeq  50$~PeV, while for $\alpha\simeq 2.5$ it would  originate from protons with $\bar{E}_p\simeq 20$~PeV. On the other hand, due to the fact that the distribution of parent proton energies extends up to very large values, see Fig.~\ref{pivsep}, the average proton energy $\langle E_p\rangle$ leading to a given value of $E_\nu$ turns out to be much higher than the median energy $\bar{E}_p$, and can be few hundred times larger than the neutrino energy if the spectrum is hard (and assuming no cutoff), as is shown in the right panel of  Fig.~\ref{emed}.

 If the sources were at cosmological distances, the observed neutrino energies would appear redshifted by a factor $(1+z)^{-1}$ with respect to the energy produced at the source\footnote{If one were to compute the cumulative flux from all such sources, it would also be necessary to integrate over the contribution from sources at all redshifts accounting for the  cosmological evolution of their density.}, but for simplicity we ignore here those effects, which anyway do not affect the ratios  $\bar{E}_p/E_i$ significantly, just shifting the energy in the horizontal axis in Fig.~\ref{emed} by that factor.

\begin{figure}[t]
\centering
\includegraphics[scale=.95,angle=0]{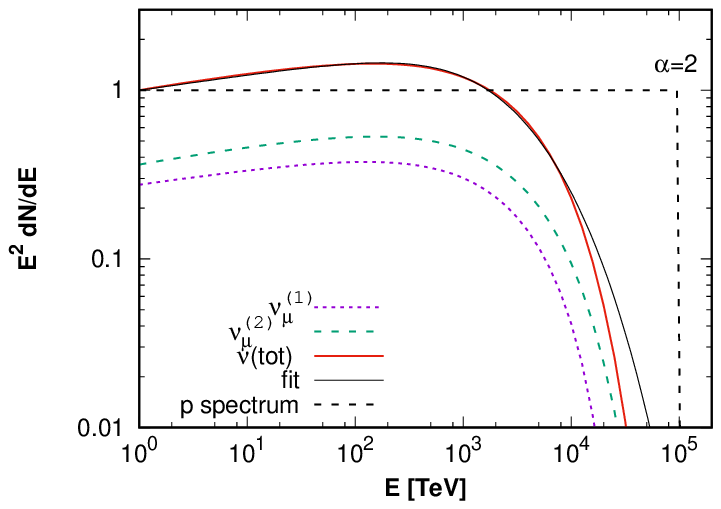}\includegraphics[scale=.95,angle=0]{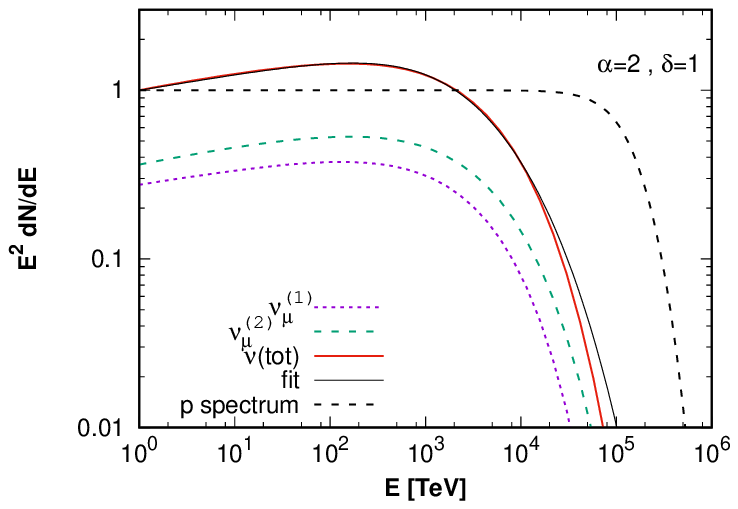}
\caption{Total $\nu$ spectrum (from all flavors), as well as the individual contributions from $\nu_\mu^{(1)}$ and $\nu_\mu^{(2)}$, from $pp$ scenarios with a power-law proton spectrum with $\alpha=2$. The left panel corresponds to a sharp cutoff at $E_c=100$~PeV while the right panel is for a smoother cutoff scaling as $\cosh^{-1}(E/E_c)$, whose shapes are shown with black dashes, arbitrarily normalized to unity at 1~TeV. Also shown is the fitting function to the total neutrino spectrum discussed in the text. }
\label{cutoff}
\end{figure}

The energy dependence of  $\bar{E}_p/E_i$, as well as the increase of the cross section and pion multiplicity with energy, also implies that the resulting neutrino spectral shape will slightly differ from that of the original proton spectrum. While the first effect would tend to make it steeper, the other two would tend to make it harder. Detailed computation shows that the neutrino spectrum tends  indeed to be slightly harder than the proton spectrum, due to the increase in the average number of neutrinos produced in $pp$ interactions with increasing energies (see e.g. \cite{ka06,ka14,su16}).
Another potential implication of these results is that they affect the relation between the cutoff energy of the CR spectrum and the associated cutoff in the neutrino spectrum.  
As an example of these effects, we show in Fig.~\ref{cutoff} the total differential neutrino flux summed over all flavors (and multiplied by $E_\nu^2$), as well as the individual contributions from $\nu_\mu^{(1)}$ and $\nu_\mu^{(2)}$, resulting from an $E^{-2}$ CR proton spectrum with a sharp cutoff at $E_c=100$~PeV (left panel) and one with a smooth cutoff of the form d$N_p/{\rm d}E\propto E^{-2}/\cosh((E/E_c)^\delta)$, with $\delta=1$ (right panel).\footnote{We use the hyperbolic cosine suppression rather than an exponential one because with it the spectrum below $E_c$ is closer to the power-law $E^{-\alpha}$ and the determination of the hardening becomes clearer.} One can see that the neutrino spectrum is harder than the original proton spectrum and it has a very smooth suppression starting at about 1~PeV. A good fit to the total neutrino spectrum can be obtained with the expression d$N_\nu/{\rm d}E\propto E^{-\gamma}/\cosh((E/E_\star)^{\delta'})$, as shown in the plot. We obtain  $\gamma\simeq 1.91$, $E_\star\simeq  E_c/107$ and $\delta'\simeq 0.45$ for the sharp cutoff case (left panel), while  $\gamma\simeq 1.90$, $E_\star\simeq  E_c/107$ and $\delta'\simeq 0.39$ for the softer cutoff case with $\delta=1$ (right panel).   Note that due to the wide spread of the  energies of the  neutrinos produced in $pp$ interactions,  one indeed expects that the falloff in the neutrino spectrum should be significantly less pronounced than that of the original CR spectrum. It is worth noting that these results cannot be obtained if one computes the neutrino production in the so-called delta function approximation, i.e. setting the pion and neutrino energies in  a $pp$ interaction to their average values (see e.g. \cite{be14}). 

Let us mention that the spectra of the different neutrinos at the source could be affected by the presence of strong (larger than 1~kG) magnetic fields in the source interaction region, since this could lead to significant synchrotron losses  of the high-energy muons before they decay, corresponding to the so-called muon damped neutrino scenarios. In this hypothetical case, above a certain critical energy the neutrino fluxes at the source 
 will be dominated by the $\nu_\mu^{(1)}$ component, with the $\nu_\mu^{(2)}$ and $\nu_e$ neutrinos from the $\mu$ decay being strongly suppressed. This critical energy can be estimated  equating the muon decay time, $t_{\rm dec}^\mu=(E_\mu/m_\mu)\tau_\mu\simeq 20 (E_\mu/{\rm PeV})$~s, with the muon synchrotron loss time, $t_{\rm sync}^\mu=(9/4)m_\mu^4/(e^4B^2E_\mu)\simeq 3\times 10^7\, {\rm s}/[(B/{\rm G})^2(E_\mu/{\rm PeV})]$, and is $E_\mu\simeq {\rm PeV}/(B/{\rm kG})$ (the associated neutrino energy would be a factor of  about 3 smaller). At energies about a factor 20 larger, the pion themselves will lose their energy by synchrotron emission before they can decay, and hence a strong suppression in the total neutrino flux will result above the corresponding neutrino  energies, with the spectrum becoming steeper by two extra powers of the energy, corresponding to the fraction of the mesons that decay before loosing significant amounts of energy, until eventually the neutrinos from kaon decays become the dominant ones \cite{rachen,wb2,ka05}. The non-observation of a strong suppression in the astrophysical neutrino fluxes at or below PeV energies hence suggests that in the $pp$ scenarios the magnetic fields in the interaction region are not much larger than a kG. One should also keep in mind that the superposition of sources with different magnetic fields, or even the emission of neutrinos from different regions of a source having different magnetic field values, could soften the suppression effect.
 
 The neutrino fluxes at the Earth are obtained after accounting for the averaged neutrino flavor oscillations, and would then have a different flavor ratio below and above the threshold associated to the muon damping. Below the muon damping threshold, the original source flavor ratio $(\nu_e:\nu_\mu:\nu_\tau)\simeq (1:2:0)$ will be converted by oscillations to an approximately equal flavor ratio $(1:1:1)$, while above the muon damping threshold (but below the pion damping regime), the source flavor ratio $(0:1:0)$ will be converted approximately to the $(0.2:0.4:0.4)$ flavor ratio.
 Note also that when the muons are damped, the ratio $\bar{E}_p/E_\nu$ will be that associated to $\nu_\mu^{(1)}$ rather than the value averaged over the different neutrino species.
 To describe in detail the neutrino spectra above the onset of the muon damping effects, one would need to consider the yield of pions from the proton interactions, obtain the yield of muons from pion decays and then  account for their energy losses before considering the neutrino production in their decays (see e.g. \cite{winter,wi12}).

For completeness, we also showed in Fig.~\ref{emed} the values of $\bar{E}_p/E_\gamma$ and  $\langle{E}_p\rangle/E_\gamma$  obtained for different gamma ray energies, which are found to differ from the usually assumed value of 10 especially for hard proton spectra.  One has to keep in mind that the high-energy photons can cascade down to lower energies either by interactions with gas or with the photons at the source (synchrotron or inverse Compton photons) or with the background of optical/infrared and microwave photons that they traverse in their trip up to us.  Hence, only for nearby sources, such as Galactic ones, that are thin to escaping photons the very-high energy gamma rays may reach the Earth unattenuated, while the weakly interacting neutrinos will always arrive unscathed.

\section{The $p\gamma$ scenarios}
In the $p\gamma$ scenarios the CR protons can interact with the dense radiation fields present in the sources, such as for instance the synchrotron photons from the co-accelerated  electrons in the AGN or GRB jets or the thermal optical/UV photons of the blue bump  produced in the inner torus near the  AGN cores. We will compute the production of secondaries in the frame in which the photon background is isotropic, denoting the quantities in this frame with a prime. In the case of production in relativistic jets, such as in GRBs or blazars, this will be the shock rest frame (SRF), and hence the energy in the comoving frame will typically appear boosted with respect to the energy $E'$ in the SRF by the Lorentz factor $\Gamma$ of the jet (and then eventually redshifted in the observer's frame, so that $E\simeq \Gamma E'/(1+z)$).

\begin{figure}[t]
\centering
\includegraphics[scale=.83,angle=0]{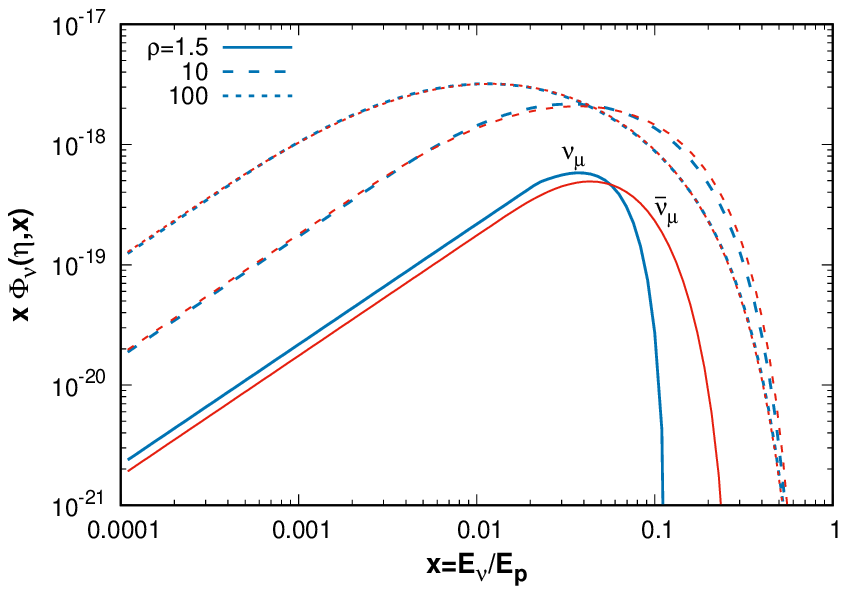}\includegraphics[scale=.83,angle=0]{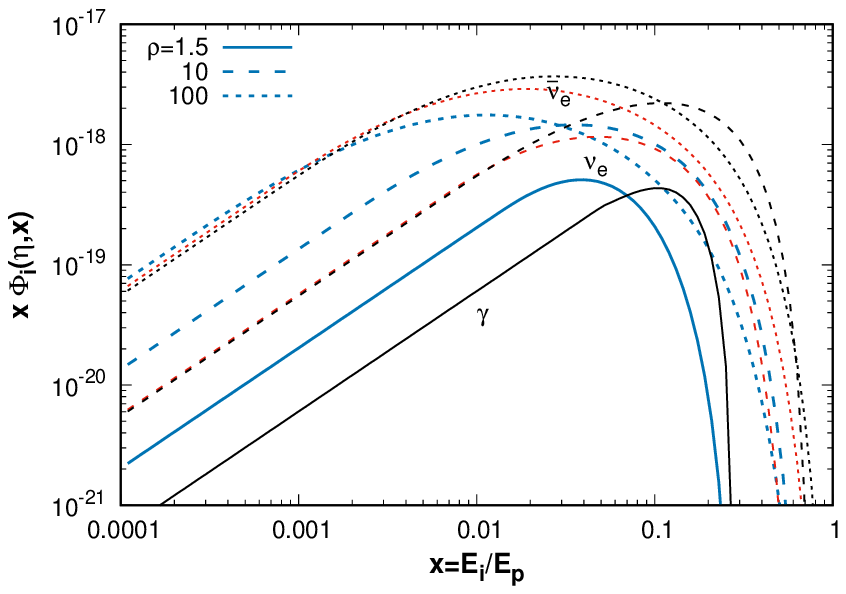}
\caption{Distribution functions of the secondaries resulting from $p\gamma$ interactions for different values of $\rho\equiv \eta/\eta_0$. The left panel is for muon neutrinos (blue) and antineutrinos (red) and the right panel for electron neutrinos (blue), electron antineutrinos (red) and photons (black). }
\label{xphivsx}
\end{figure}

The neutrino (and $\gamma$) fluxes from the decays of mesons produced in $p\gamma$ interactions can be computed as \cite{ke08}
\begin{equation}
    \frac{{\rm d}N_{i}}{{\rm d}E'_i}(E'_i)=\int \frac{{\rm d}E'_p}{E'_p}\,{\rm d}\varepsilon' \, f_p(E'_p)f_{\rm ph}(\varepsilon')\Phi_{i}(\eta,x_i),
\end{equation}
where $\varepsilon'$ is the energy of the isotropic target photons, with $f_{\rm ph}(\varepsilon'){\rm d}\varepsilon'$ and $f_p(E'_p){\rm d}E'_p$  being the number densities of background photons and CR protons in the energy intervals d$\varepsilon'$ and d$E'_p$ respectively. The variable $\eta\equiv 4\varepsilon' E'_p/m_p^2$ characterizes the center of mass total energy squared of the interaction and $x_i=E'_i/E'_p$ is the fraction of the proton energy carried by the secondary, with $i$ describing the (anti) neutrino type or the photons. 
The threshold condition to produce a pion is that $(p_\gamma+p_p)^2>(m_p+m_\pi)^2$, and one then has that
\begin{equation}
    \eta>\frac{2m_\pi}{m_p}+\frac{m_\pi^2}{m_p^2}\equiv \eta_0\simeq 0.313,
\end{equation}
where we considered for simplicity $m_{\pi^{+}}\simeq m_{\pi^0}\simeq 137$~MeV. Hence, for a given photon energy $\varepsilon'$ this requires that $E'_p>70\,{\rm TeV}/(\varepsilon'/{\rm keV})$.
The different functions $\Phi_i(\eta,x_i)$ where obtained in \cite{ke08} using SOPHIA based \cite{sophia} simulations and were conveniently parameterized there, with the parameters of the fits being tabulated for different values of $\eta/\eta_0$.\footnote{We used a linear interpolation between the tabulated values, and for $\eta/\eta_0>100$ we used for it the values of $\Phi_i(100\eta_0,x)$ in order to be able to approximately extend the computation up to the highest energies that we considered.} These functions are plotted in Fig.~\ref{xphivsx} for the different neutrino and antineutrino flavors as well as for the photons (although  additional photons may result for instance from cascading of secondary electrons).

Near the threshold of the $p\gamma$ interaction, the pion production is dominated by the $\Delta(1232)$ resonance (leading to a ratio $\pi^0/\pi^{+}\simeq 2$  because of isospin, or equivalently because $\pi^0$ can result from the production of a $u\bar u$ or a $d\bar d$ pair  while $\pi^{+}$ only from  $d\bar d$ pair production), with also some contribution to $\pi^{+}$ production from direct $t$-channel $\pi^{+}$ exchange (since a charged particle is required to couple to the photon), and then other resonances (such as $\Delta'$, $\rho$, etc.) contribute above the $\Delta$ resonance. For $\eta>2.14\eta_0$ the multipion production becomes possible and it actually becomes the dominant channel for $\sqrt{s}\gg {\rm GeV}$, leading to comparable amounts of $\pi^{+}$, $\pi^0$ and $\pi^{-}$. Other mesons get also produced, but in smaller amounts, with the neutrino production from $K^\pm$ decays being only potentially relevant at very high energies in scenarios in which the magnetic fields at the source are so strong that the charged pions get damped before they can decay. Note that for beamed emission from relativistic jets, the condition for the muons to get damped before decaying translates into $E_\mu\simeq (\Gamma/100)\,{\rm PeV}/(B'/100\,{\rm kG})$, with $B'$ being the magnetic field in the SRF (where $\Gamma=O(10)$ for AGN jets, while it is $O(10^2$--$10^3$) for GRB jets).
Before the onset of multipion production, no $\pi^{-}$ get produced in $p\gamma$ interactions, and hence fluxes of ${\bar\nu}_e$ appear at the source only above the threshold for two pion production, i.e. for $\eta>2.14\eta_0$ (we neglect the $\bar\nu_e$ from the decays of the produced neutrons, since their energies are very small). One should keep in mind however that a flux of $\bar\nu_e$ will anyhow result at the Earth from the oscillations of the $\bar\nu_\mu$ produced at the source.

We will consider in the following two specific $p\gamma$ scenarios to study their characteristic features. One is a typical GRB model, in which the target photon spectrum consists of a broken power-law
with a break at $\varepsilon'_{\rm b}=1$~keV (corresponding to the observed GRB spectral breaks at photon energies of few hundred keV), with $f_{\rm ph}=N_\gamma (\varepsilon'/\varepsilon'_{\rm b})^{-\beta}$, where $\beta=1$ for $\varepsilon'<\varepsilon'_{\rm b}$  and $\beta=2$ for $\varepsilon'>\varepsilon'_{\rm b}$, and considering photon energies in the range  $2\,{\rm eV}<\varepsilon'<300$~keV. The other is a thermal photon spectrum with $kT=10$~eV, typical of the AGN's blue bump.\footnote{One has to keep in mind however that in this model  the photon source  would be external to the jet where CRs get accelerated, and hence in reality would not be exactly isotropic in the interaction region.}

\begin{figure}[t]
\centering
\includegraphics[scale=.83,angle=0]{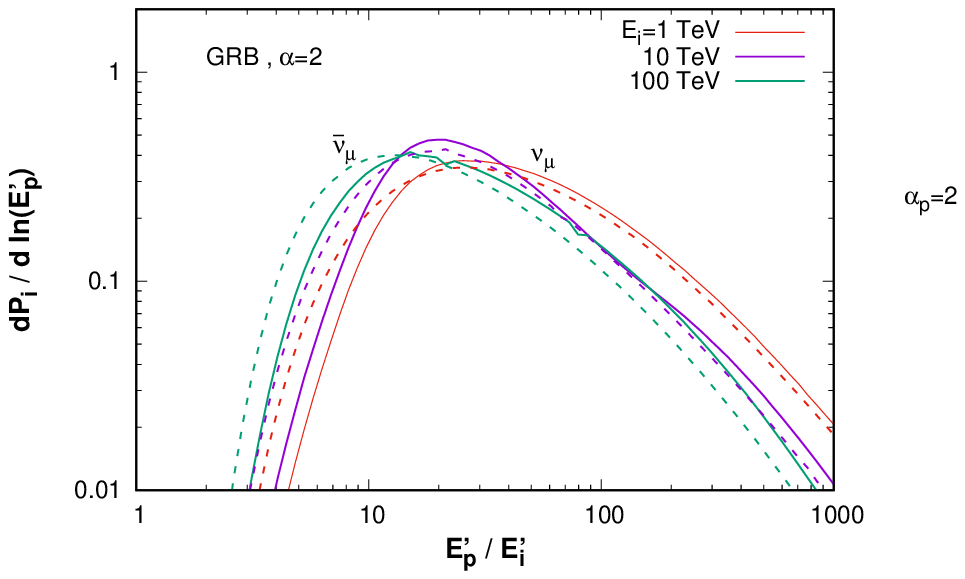}\includegraphics[scale=.83,angle=0]{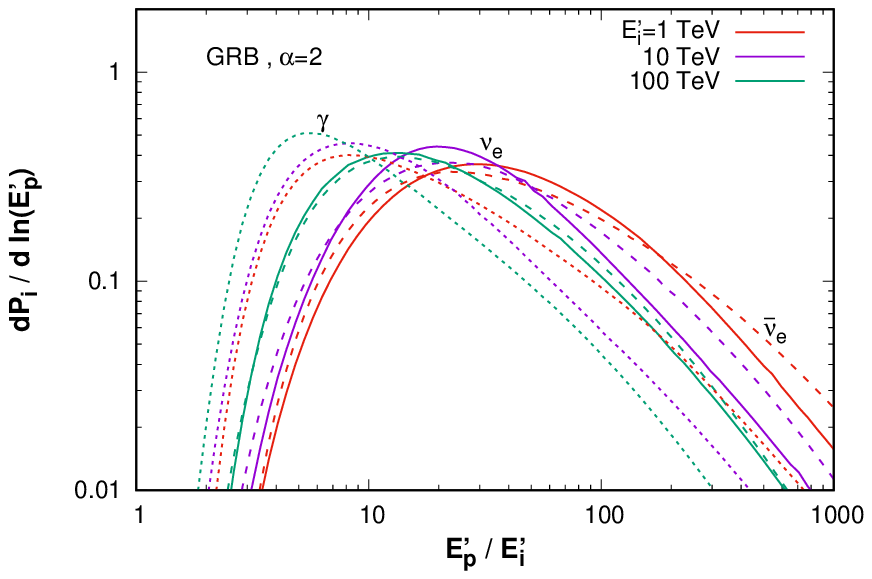}
\caption{Probability distribution for the parent proton energies that produce neutrinos or photons with different energies in GRB scenarios, adopting $\alpha= 2$. }
\label{pgrb}
\end{figure}

\begin{figure}[t]
\centering
\includegraphics[scale=.95,angle=0]{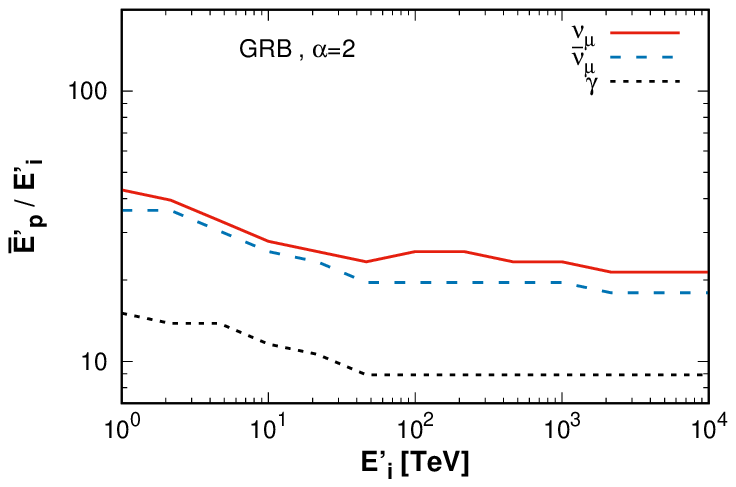}\includegraphics[scale=.95,angle=0]{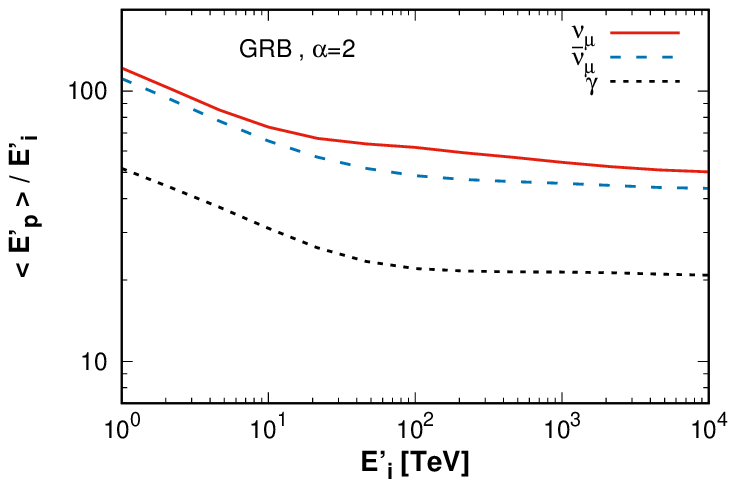}
\caption{Left panel: ratio $\bar E'_p/E'_i$  as a function of $E'_i$ in  $p\gamma$ GRB scenarios, adopting  $\alpha=2$. Right panel: the same in terms of the average proton energy $\langle E_p'\rangle$.}
\label{emedgrb}
\end{figure}

Considering first the GRB scenario, we show in Fig.~\ref{pgrb} the probability distribution of the proton energies giving rise to different neutrino energies for the different flavors. The left panel is for  $\nu_\mu$ and $\bar{\nu}_\mu$, while the right panel is for $\nu_e$, $\bar\nu_e$ and $\gamma$. In the left-panel of Fig.~\ref{emedgrb} we show the median proton energies $\bar{E}_p$, normalized to the corresponding neutrino (or photon) energies $E_i$, as a function of the neutrino (or $\gamma$) energies. The main distinguishing feature observed is the rise in the ratio $\bar{E}_p/E_i$ below $E'_i\simeq 100$~TeV, due to the fact that in this case the threshold condition for $\Delta$ production requires median proton energies  in excess of $20E_i$ (see also the shift in the distributions in  Fig.~\ref{pgrb} for decreasing $E_i$).  The results for the electron flavor are not very different from those shown for $\bar\nu_\mu$. The right panel is similar but in terms of the average energy  of the protons $\langle E'_p\rangle$ that give rise to a given secondary energy $E'_i$, in which case the effect is much more pronounced.

\begin{figure}[t]
\centering
\includegraphics[scale=.95,angle=0]{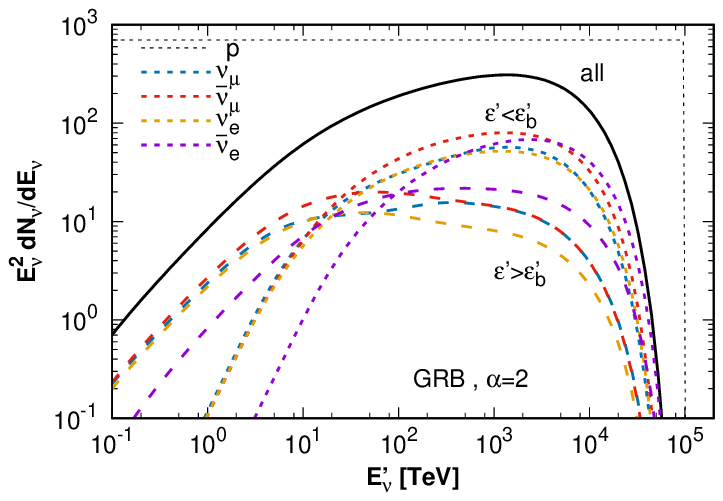}\includegraphics[scale=.95,angle=0]{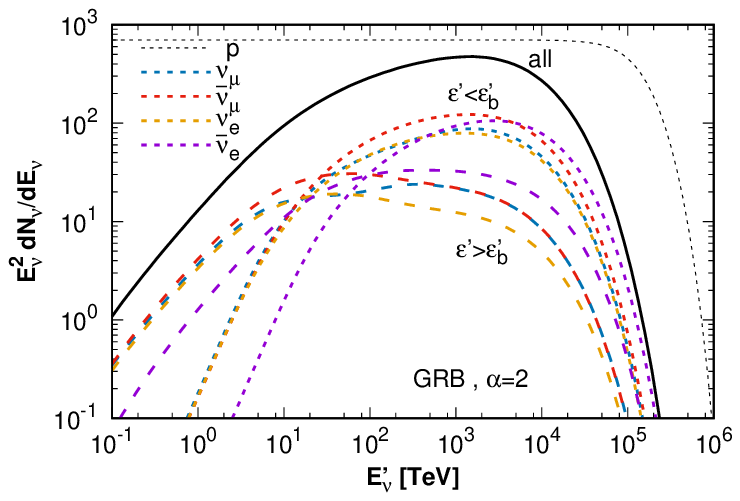}
\caption{Total all flavor $\nu$ spectrum in GRB scenarios (solid line), adopting a power-law proton spectrum with $\alpha=2$ having a sharp cutoff at 100~PeV (left panel) or a smooth one with $\delta=1$ (right panel).  Also shown are the different flavor contributions (at the source) arising from the interactions with the photons with $\varepsilon'<\varepsilon'_{\rm b}$ and with $\varepsilon'>\varepsilon'_{\rm b}$. Black dashed lines illustrate the shape of the proton spectrum adopted (with arbitrary normalization).}
\label{spectgrb}
\end{figure}

We then show in Fig.~\ref{spectgrb} the resulting neutrino spectrum in the SRF, for an assumed power-law proton spectrum with $\alpha=2$ and a sharp cutoff at $E'_p=100$~PeV (left panel) or a smooth one with the hyperbolic cosine suppression with $\delta=1$ (right panel). In order to better understand the origin of the neutrinos with different energies, we also show the separate contributions to the flux arising from the interactions with photons with $\varepsilon'<\varepsilon'_{\rm b}$ and $\varepsilon'>\varepsilon'_{\rm b}$, for the different flavors\footnote{Note that SOPHIA does not account for the polarization of the decaying muons, but the associated effects are small in comparison with the uncertainties on the assumed photon spectra, the CR spectral cutoff shape, etc.}. Most of the high-energy neutrinos in the flatter region above 10~TeV turn out to originate from interactions with the photons with the harder spectrum present below $\varepsilon'_{\rm b}$, and actually the bulk of the contribution arises from photons with energies within an order of magnitude of $\varepsilon'_{\rm b}$. The spectral cutoff in the neutrino spectrum is found to be much softer than the cutoff adopted for the CR spectrum. We also note that often the non-thermal radiation in AGN jets is modelled as a photon background scaling as $\varepsilon'^{-1}$ which extends up to about 1~keV, and hence in this case the resulting neutrino spectrum would be similar to the GRB component computed for $\varepsilon'<\varepsilon'_{\rm b}$ that is shown in the figure.

\begin{figure}[t]
\centering
\includegraphics[scale=.83,angle=0]{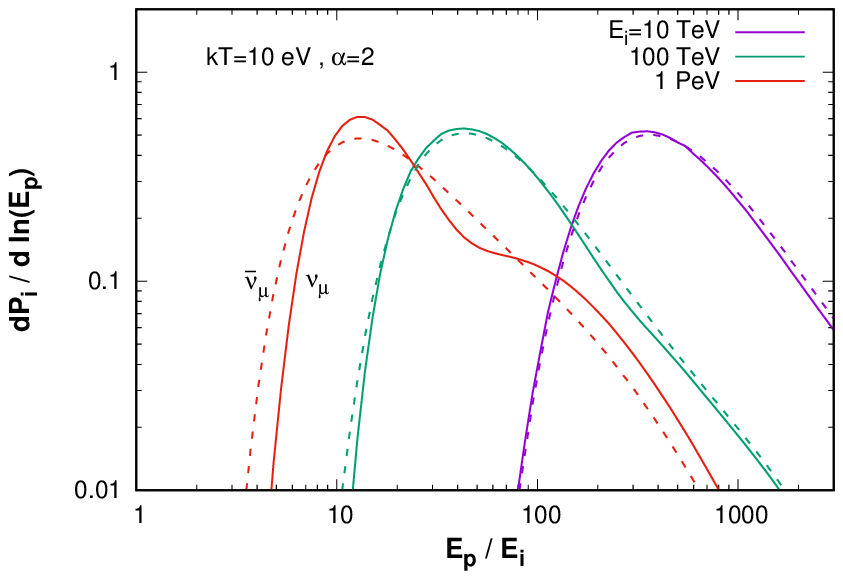}\includegraphics[scale=.83,angle=0]{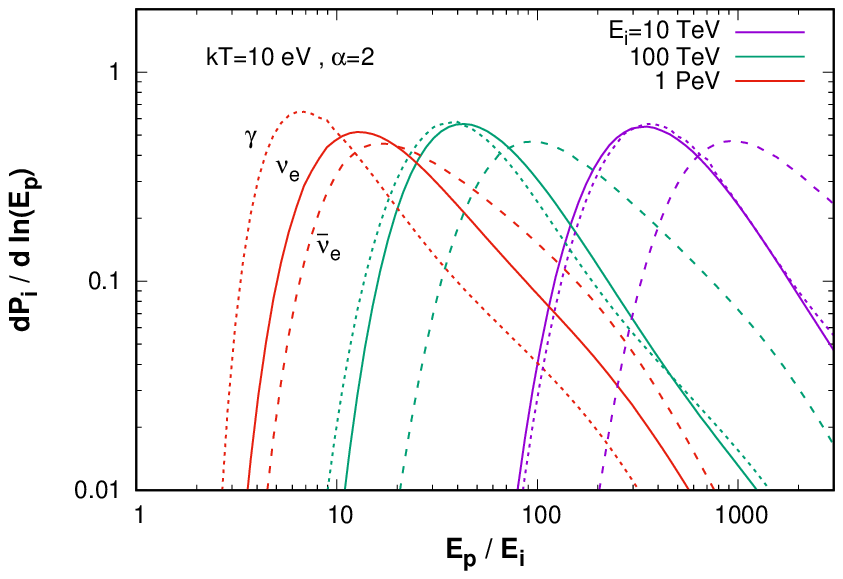}
\caption{Probability distribution for the parent proton energies that produce neutrinos or photons with different energies $E_i$ in scenarios with a thermal photon background with $kT=10$~eV, adopting $\alpha= 2$. }
\label{pvszth}
\end{figure}

\begin{figure}[t]
\centering
\includegraphics[scale=.97,angle=0]{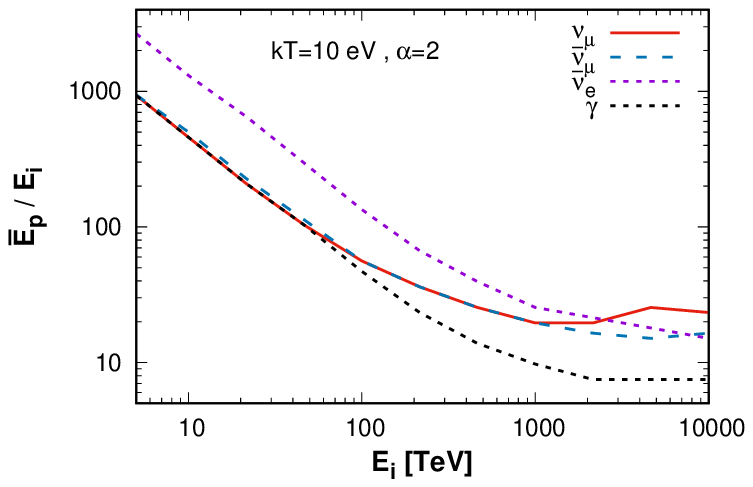}\includegraphics[scale=.97,angle=0]{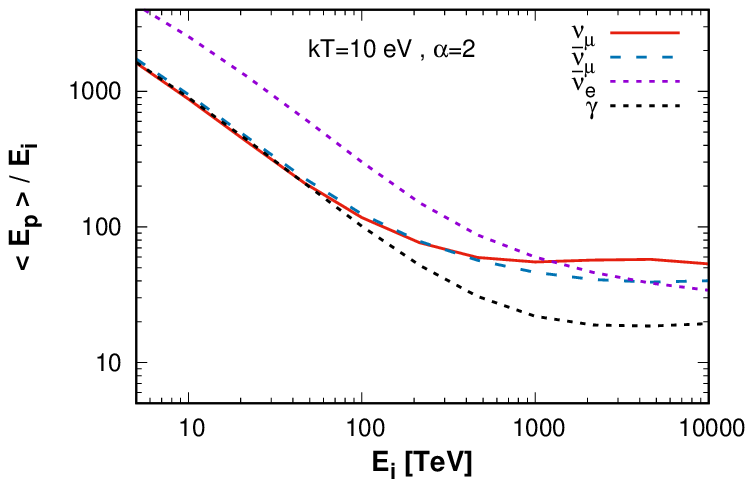}
\caption{Left panel: ratio $\bar E_p/E_i$  as a function of $E_i$ in  $p\gamma$ scenarios with a thermal photon background with $kT=10$~eV, adopting  $\alpha=2$. The results for $\nu_e$ are comparable to those of $\bar\nu_\mu$. Right panel: the same in terms of the average proton energy $\langle E_p\rangle$.}
\label{z05vseith}
\end{figure}

\begin{figure}[h]
\centering
\includegraphics[scale=.95,angle=0]{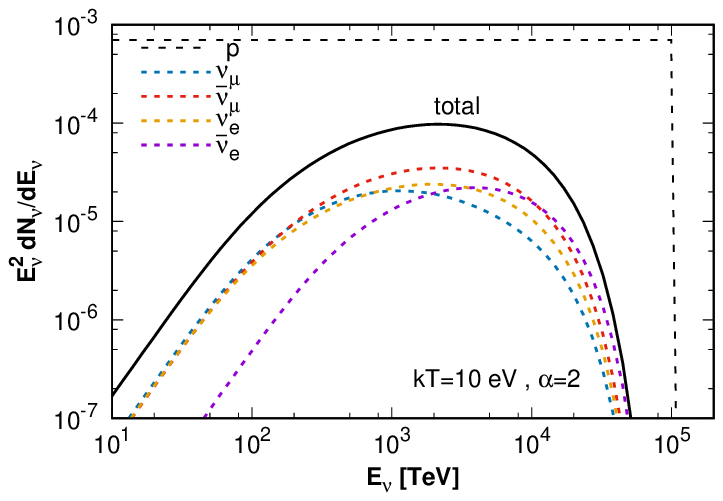}\includegraphics[scale=.95,angle=0]{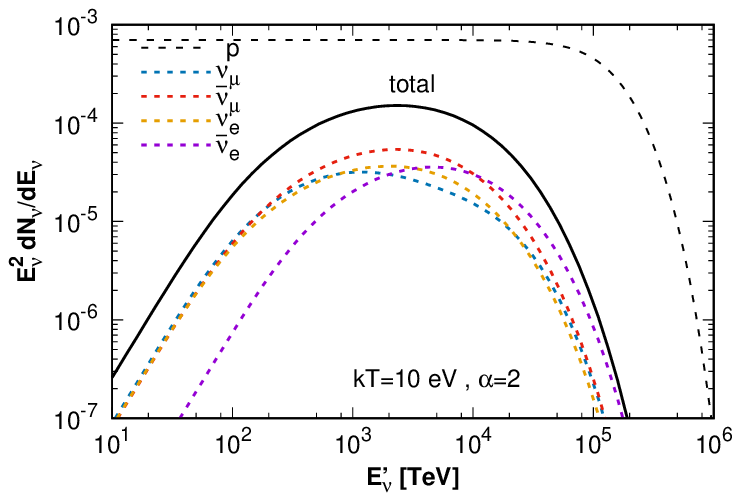}
\caption{Total source $\nu$ spectrum as well as the different flavor contributions in $p\gamma$ scenarios with a thermal photon background with $kT=10$~eV, adopting a power-law proton spectrum with $\alpha=2$ having a sharp cutoff at 100~PeV (left panel) or a smooth one with $\delta=1$ (right panel).  Black dashed lines illustrate the shape of the proton spectrum adopted (with arbitrary normalization).}
\label{nuspecth}
\end{figure}

Considering now the case of a thermal radiation background with $kT=10$~eV, characteristic of the AGN blue bump, we show in Fig.~\ref{pvszth} the distribution of probabilities for the different neutrino flavors and for the gamma rays. In Fig.~\ref{z05vseith} we show the corresponding median and average proton energies giving rise to secondaries with energies $E_i$. The main difference with the GRB example is that, given that the photon energies are on average much smaller and that they have a narrower distribution, as a consequence of the effect of the threshold for $\Delta$ production the required proton energies to give rise to a given neutrino energy can be much larger than in the previous example. In particular, the pronounced increase in $\bar{E}_p/E_i$ for  values of $E_i$ smaller than few hundred TeV can be explained because the pions are produced mostly at the $\Delta$ resonance and the main contribution to the neutrinos comes from the low energy tail of the distribution of decay products. One can see that in this case the median energy of the protons remains at a value of about 5~PeV for neutrinos with energies below 200~TeV (for the $\bar\nu_e$ the median proton energy is a factor about 2.5 times higher, since their production requires the multipion production to be allowed).  Since the low energy tail of the neutrino distribution is quite suppressed, one then expects a strong suppression to appear in the neutrino spectrum at low energies. The resulting spectrum is shown in Fig.~\ref{nuspecth}, where indeed one can see that it steeply  increases with energy and only flattens at energies above few hundred TeV. The spectral cutoff in the neutrino spectrum is also found to be much smoother than the cutoff adopted for the CR spectrum.

\section{Discussion}

We have studied the distribution of the energies of the CR protons giving rise to astrophysical neutrinos with a given energy, both for the $pp$ and the $p\gamma$ scenarios. In the $pp$ scenarios, given the low energy threshold for pion production and the slow (logarithmic) rise in the inelastic cross section and charged particle multiplicities, for power-law CR spectra the neutrino spectral shape is quite similar, though slightly harder. On the other hand, for increasing energies the median energy of the protons giving rise to neutrinos steadily increases beyond the usual estimate of $E_p\simeq 20E_\nu$, and depends on the production channel (pion or muon decay).
 This increase depends on the CR spectral slope, and for instance it is about $\bar{E}_p\simeq 50E_\nu$ for $\alpha=2$, becoming  smaller for softer CR spectra.  This has an impact also  on the relationship between the CR spectral cutoff and the resulting neutrino cutoff, which becomes much smoother than the proton one  once the broad distribution of the energies of the produced neutrinos is taken into account. 
  In this sense, if a sharp suppression in the  astrophysical neutrino  spectrum were to be observed, this could be an indication in favor of the presence of strong magnetic fields in the interaction region that damp the decaying charged mesons as a consequence of their synchrotron losses. However, for the effect to be sharp the magnetic field values in the neutrino producing regions would be required to be similar in most of the sources. On the other hand, the observation of an astrophysical neutrino spectrum with an approximately power-law shape could be an indication of an underlying $pp$ scenario with a large proton cutoff energy. A distinguishing feature of the $pp$ scenarios is that the spectrum of both the generated neutrinos and the photons keep their approximate power law shape down to lower energies, and through the cascading of the photons by interactions with the EBL radiation this could overshoot the observed diffuse gamma ray fluxes below TeV energies if the CR spectrum is soft ($\alpha\geq 2.2$) \cite{mu13,be17}. This tension may be overcome for hidden neutrino sources, i.e. those that  are opaque to the photons.
 
In the $p\gamma$ scenarios, the threshold for the $\Delta$ production has a strong impact on the neutrino production, leading to a steep suppression of the neutrino fluxes below an energy $E_\nu\sim 70\,{\rm TeV}/(\varepsilon/{\rm keV})$ for target photons with energy $\varepsilon$.  This also leads to an increase in the values of $\bar{E}_p/E_\nu$ for lower energies, and in particular in the case considered of the relatively narrow thermal distribution with $kT=10$~eV we found that due to the effect of the $\Delta$-resonance threshold  the value of $\bar{E}_p$ remains about 5~PeV for $E_\nu<200$~TeV (being about 12~PeV for the case of $\bar\nu_e$).  For target photons with a broken power-law energy distribution, as the one that we considered with a break  at $\varepsilon'_{\rm b}=1$~keV characteristic of GRB scenarios, the high energy neutrinos with $E'_\nu>10$~TeV get produced predominantly through interactions with the photons  below the break, which have a harder spectrum, but due to the much broader photon spectral distribution the changes in the values  of $\bar{E}_p/E_\nu$ are less pronounced than in the case of a thermal photon distribution.

To conclude,  the research field of high-energy neutrinos is today in a very exciting state with great promise for the future, particularly in terms of what it will teach us about the sources of cosmic rays. We have quantitatively discussed the connection between neutrinos and cosmic rays, examining the correspondence between the energies of the primaries and of the secondaries for different source scenarios. Knowing which kind of sources produce the neutrinos will be fundamental for a proper interpretation of the results and  the identification of specific features in the neutrino spectrum will be of crucial importance for this.

\section*{Acknowledgments}
The work of E.R. was supported by CONICET (PIP 2015-0369) and ANPCyT (PICT 2016-0660). The work of F.V. was partially supported by the grant 2017W4HA7S ``NAT-NET: Neutrino and Astroparticle Theory Network'' under the program PRIN 2017 funded by the Italian Ministero dell'Istruzione, dell'Universit\`a e della Ricerca (MIUR).


\end{document}